\documentclass[aps,prl,reprint,groupedaddress,longbibliography]{revtex4-2}
\usepackage{graphicx}
\usepackage{bm}
\usepackage{amsmath}
\usepackage{braket}

\graphicspath{{./},{./figs/}}

\begin{document}

\title{Relation between dynamic heterogeneities observed in scattering experiments and four-body correlations}

\author{Kohsei Kanayama$^1$}
\author{Taiki Hoshino$^2$}
\author{Ryoichi Yamamoto$^1$}
\email{ryoichi@cheme.kyoto-u.ac.jp}
\affiliation{
$^1$Department of Chemical Engineering, Kyoto University, Kyoto 615-8510, Japan\\
$^2$RIKEN SPring-8 Center, 1-1-1 Kouto, Sayo-cho, Sayo-gun, Hyogo 679-5148, Japan
}

\date{\today}

\begin{abstract}
{\it Dynamic heterogeneity} is expected to be a key concept for understanding the origin of slow dynamics near the glass transition. In previous studies, quantitative evaluations of dynamic heterogeneity have been attempted using two different routes, {\it i.e.}, the speckle patterns in scattering experiments or the four-body correlation functions of microscopic configuration data obtained from molecular dynamics simulations or real-space observations using confocal microscopy. However, the physical relationship between these dynamic heterogeneities obtained using different methods has not been clarified. This study proposes a connection between dynamic heterogeneities characterized based on speckle patterns and those obtained from four-body correlations. The validity of the relationship is also clarified.
\end{abstract}

\maketitle


The glass transition is a common phenomenon observed in metals, polymers, molecular and ionic liquids, and colloidal dispersions \cite{angell1995, berthier2011glass}.
However, this phenomenon has remained a mystery in many areas for decades, such as the dramatically slow dynamics near the glass transition upon cooling; consequently, it is recognized as an open problem in physics.
As an example of this problem, a dramatic increase in the viscosity of a liquid near its glass transition temperature is experimentally observed.
One suggestion for the cause of this dynamic behavior is the cooperative movement of the molecules in a supercooled liquid.
{\it Dynamic heterogeneity} (DH) has been proposed in previous studies \cite{ediger2000, berthier2011} to understand the mechanism underlying these molecular dynamics.
This concept characterizes the heterogeneity of molecules' mobility dependent on their positions in a liquid.
The idea of DH is helpful for understanding the origin of the slow dynamics near the glass transition.

Two methods for evaluating DH have been reported to date to quantify the phenomenon of the cooperative movement of molecules;
one is based on a four-body correlation function $S_4$ (Approach 1), and the other is based on the scattering intensity $I$ of the speckle patterns measured in a scattering experiment (Approach 2).

In Approach 1, one calculates the function $S_4$ from microscopic configuration data generated in molecular dynamics (MD) simulations or obtained from real-space observations using confocal microscopy. Many studies on this topic have been reported since approximately 2000 \cite{Mizuno2012, glotzer2000, doliwa2000, lavcevic2003, berthier2004, toninelli2005, Chandler2006, stein2008, karmakar2010, flenner2010, flenner2011}, where several observables were used to measure the mobility of each particle to evaluate the DH.
However, this method is not applicable in experiments because the information of each individual particle cannot be measured.
Therefore, researchers have not yet clearly determined whether the results obtained using Approach 1 are valid in reality due to a lack of any method for confirmation.
On the other hand, in Approach 2, one evaluates DH based on the scattering intensity $I$ measured in a scattering experiment, such as X-ray photon correlation spectroscopy (XPCS).
According to previous studies \cite{mayer2004, duri2005, berthier2011, hoshino2020}, DH is quantitatively evaluated using the scattering intensity of speckle patterns measured in this type of experiment.
Using this method, one can determine the actual extent of the heterogeneity of the dynamics. This approach has also been applied to computer simulations, such as MD simulations.
However, to our knowledge, no reports have directly compared the two approaches yet after some earlier related studies, found for example in \cite{Berthier2007}.

Based on the background described above, we have studied the relation between Approaches 1 and 2.
The relation between these two approaches must be determined because they may focus on different physical origins.
We applied the two approaches individually to the same system via MD simulations and evaluated the DH intensity, representing the amplitude of the spatio-temporal variations of slow/fast domains, using each approach independently to investigate this relation.
We examined the dependence of the extent to which a state is supercooled on the intensity by setting different temperatures, representing states ranging from a normal liquid to a highly supercooled liquid.
In this paper, we will present the two types of DH intensity calculated using Approaches 1 and 2 and discuss the relation between them by comparing the corresponding results.

This paper is organized as described below.
After our MD simulation model is briefly explained, we quantitatively show the methods we used to evaluate DH with this model.
The results of the two evaluation approaches introduced above, {\it i.e.}, by calculating the four-body correlation function and using the scattering intensity, are also presented.
Finally, we discuss the relation between the two approaches by comparing their results and summarize our findings.

We have conducted MD simulations in three dimensions to investigate the origin of the slow dynamics in a supercooled state.
Our simulation model is composed of two types of particles, 1 and 2, whose sizes and masses differ.
The mass ratio between these two types of particles is set to $m_2/m_1=2$, and the size ratio is set to $\sigma_2/\sigma_1=1.2$, which is effective for preventing crystallization of the system at low temperatures \cite{Miyagawa1991}.
In this model, a repulsive soft-sphere potential exists between particles:
\begin{equation}
    v_{\alpha\beta}(r) = \epsilon(\sigma_{\alpha\beta} / r)^{12}, \,\,\,\,\,\, 
    \sigma_{\alpha\beta} = ({\sigma_\alpha}+{\sigma_\beta})/2,
    \label{eq: potential}
\end{equation}
where $r$ is the distance between two particles and $\sigma_{\alpha}$ and $\sigma_{\beta}$ are the radii of particles $\alpha,\beta \in 1, 2$, respectively.
$\epsilon$ represents the strength of the pair-interaction, which is truncated at $r=3\sigma_{\alpha\beta}$.
Spatial distance, time and temperature are measured in units of $\sigma_1$, $\tau_0=(m_1\sigma_1^2/\epsilon)^{1/2}$ and $\epsilon/k_{\rm B}$, respectively.
We set the number of particles in the whole system to $N=1\times10^5$, which includes equal numbers of particles of each type, {\it i.e.}, $N_1=N_2=5\times10^4$,
and we fix the density to $\rho=N/V=0.8/\sigma_1^3$.
Then, the system length is $L=V^{1/3}=50.0~\sigma_1$.
We set different temperatures of $k_{\rm B}T/\epsilon=0.772$, $0.473$, $0.352$, $0.306$, and $0.267$ to assess the dependence of DH on the degree of supercooling of the system.
Note that the freezing point of the corresponding binary mixture is approximately $T=0.772$ \cite{Miyagawa1991},
below which the system is in a supercooled state.

\begin{figure}[tbp] 
\begin{minipage}[b]{1.0\linewidth}
        \centering
        \includegraphics[clip, width=1.0\linewidth]{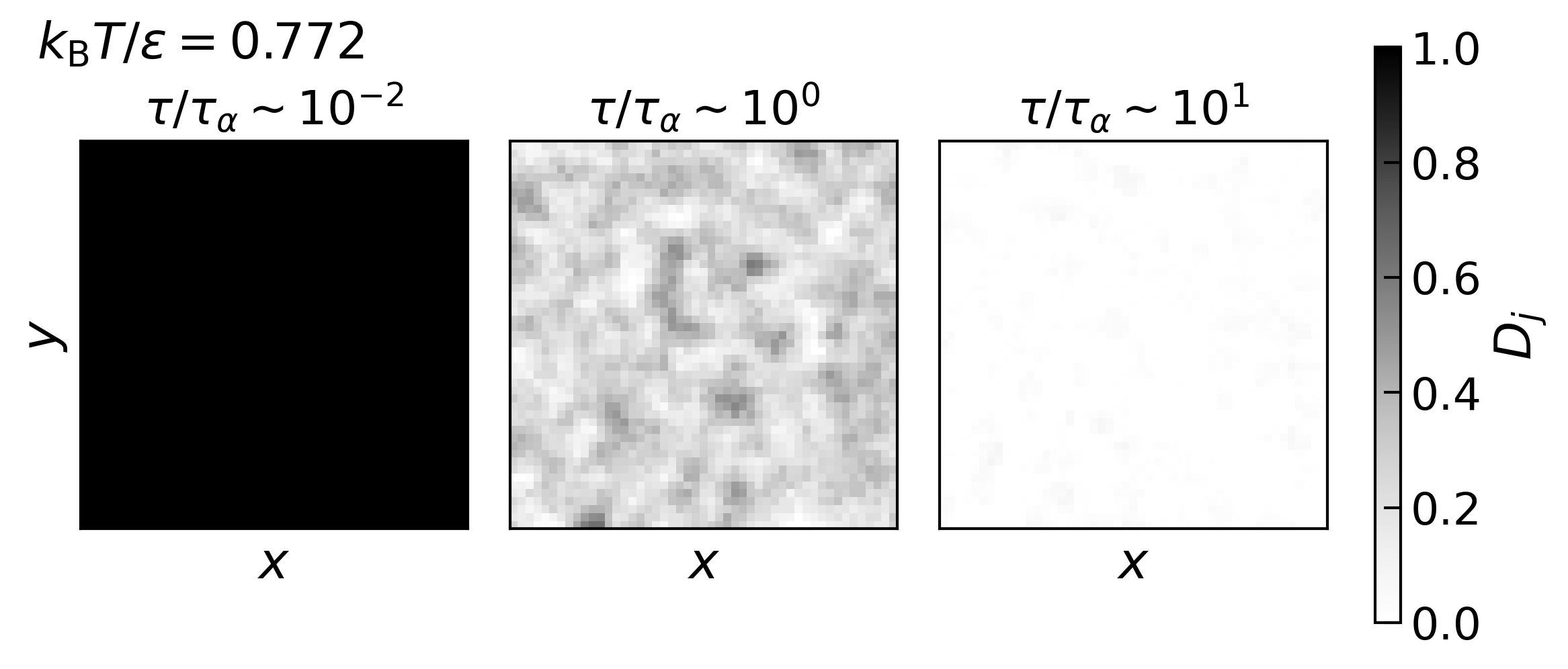}
\end{minipage}
\begin{minipage}[b]{1.0\linewidth}
        \centering
        \includegraphics[clip, width=1.0\linewidth]{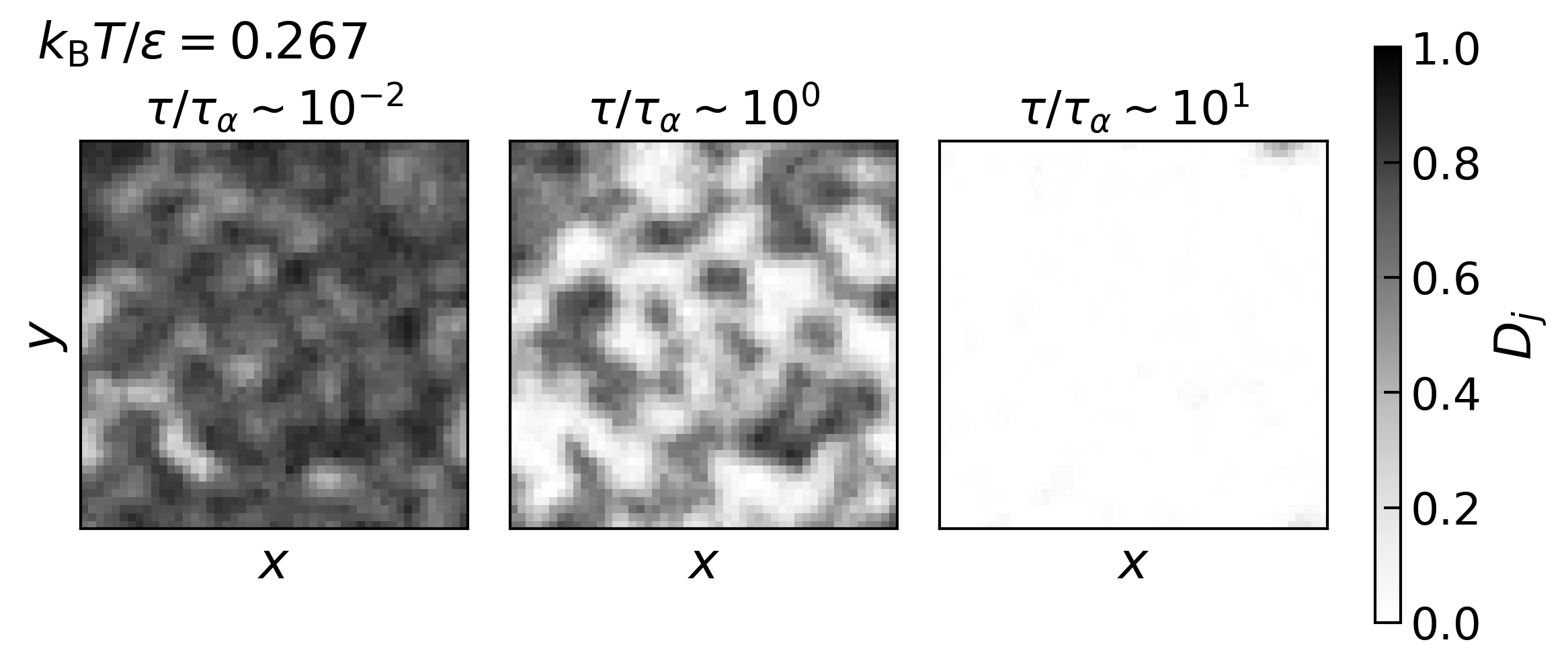}
\end{minipage}\\
\caption{
The spatial distributions of $D_{j}$ are drawn on an $xy$ cross-sectional plane of the 3D system as a function of the time interval $\tau$ from a fixed initial time $t$.
The upper panels correspond to $k_{\rm B}T/\epsilon=0.772$, and the lower panels correspond to $k_{\rm B}T/\epsilon=0.267$.
According to Eq. (\ref{eq: particle_mobility}), the value of $D_{j}$ changes from $1$ to $0$ as the particle displacements increase with time interval $\tau$ (increasing from left to right).
Although $D_{j}$ uniformly is assigned a value of $1$ (black) at $\tau=0$ or $0$ (white) for $\tau\gg\tau_\alpha$, notable heterogeneity appears at intermediate time intervals $\tau\simeq \tau_\alpha$, where darker (or lighter) colors represent the heterogeneous domains in which the dynamics of the particles are faster (or slower) than the average.
Clearly, the variance (intensity) of the heterogeneity is larger at $k_{\rm B}T/\epsilon=0.267$ than at $0.772$ around $\tau\simeq \tau_\alpha$.
}
    \label{fig: order_param}
\end{figure}

In previous studies \cite{Mizuno2012}, the intensity of DH has been quantitatively evaluated by calculating the four-body correlation function $S_{4,k}$.
In this approach, one obtains $S_{4,k}$ from an order parameter $Q_k$, which indicates the mobility (or immobility) of each particle in the system.
Several forms of the order parameter have been suggested \cite{Mizuno2012, glotzer2000, doliwa2000, lavcevic2003, berthier2004, toninelli2005, Chandler2006, stein2008, karmakar2010, flenner2010, flenner2011}.
In this study, we define it as
\begin{equation}
    Q_k(\boldsymbol{r},t,\tau) = \frac{1}{N\left<{\sigma_j^3}\right>} \sum^{N}_{j=1} {\delta}D_j \sigma_j^3 \delta\left[\boldsymbol{r}-\boldsymbol{r}_j(t)\right] ,
    \label{eq: order_param}
\end{equation}
\begin{equation}
    \hat{Q}_k(\boldsymbol{q},t,\tau) = \frac{1}{N\left<{\sigma_j^3}\right>} \sum^{N}_{j=1} {\delta}D_j\sigma_j^3 e^{ -i \bm{q} \cdot \bm{r}_j(t) },
    \label{eq: order_param_FT}
\end{equation}
where $\bm{r}_j$ is the position vector of particle $j$ in real space and the bracket $\langle{\cdots}\rangle$ denotes the ensemble average over all particles.
The hat $\hat{\ }$ denotes a value in Fourier space, and $\bm{q}$ is the wave vector corresponding to the position vector $\bm{r}$.
In this order parameter, the particle mobility deviation ${\delta}D_j$ is weighted by each particle position considering each volume ratio $\sigma_j^3/\langle{\sigma_j^3}\rangle$.
We quantified the mobility deviation by defining the {\it immobility} of particle as
\begin{equation}
    \begin{split}
        D_j&=D_j(k_m,t,\tau) = \left<{ e^{-i\boldsymbol{k}\cdot \Delta\boldsymbol{r}_j(t,\tau)} }\right>_{k_m}, \\
        {\delta}D_j &= \langle D \rangle_t - D_j,\ \ {\rm where}\ \ D\equiv{ \frac{1}{N} \sum_{j=1}^N{D_j} }.
    \end{split} 
    \label{eq: particle_mobility}
\end{equation}
Here, $\Delta\boldsymbol{r}_j(t,\tau)$ is the displacement of particle $j$ between ${\rm time}=t$ and $t+\tau$.
$\bm{k}$ represents the wave vector corresponding to the particle displacement $\Delta\boldsymbol{r}$.
The average $\langle{\cdots}\rangle_{k_m}$ is calculated for all wave vectors $\bm{k}$ that are consistent with the periodic boundary condition and satisfy $k_m-\delta k_m\le|\bm{k}|\le k_m+\delta k_m$, with $k_m\equiv 2\pi$ and $\delta k_m=0.001k_m$ for Eq. (\ref{eq: particle_mobility}) or $0.01k_m$ for Eq. (\ref{eq: two-time_corrfunc}), assuming isotropy of the system.
$D_j$ is defined as any quantity that represents the mobility (or immobility) of particle $j$ from time $t$ to $t+\tau$. 
However, in this study, we use the definition in Eq.~(\ref{eq: particle_mobility}) for better consistency with the latter method that uses the scattering intensity. 
In this definition, the value of $D_j$ changes from $1$ to $0$ when the displacement of the particle $|\Delta\boldsymbol{r}_j(t,\tau)|$ becomes comparable to its radius in the time interval $\tau$.
The mean value averaged over time $t$ is presented as $\langle{\cdots}\rangle_{t}$.
Therefore, ${\delta}D_j$ is the deviation of the mobility of particle $j$ from the mean value.
We provide the reader a more intuitive understanding of the order parameter $Q_k(\bm{r}, t, \tau)$ by illustrating 
the spatial distributions of $D_{j}$ at three time intervals, $\tau/\tau_\alpha=10^{-2}$, $1$, and $10$, 
from a fixed initial time $t$ in Fig.~\ref{fig: order_param}
for a normal liquid ($k_{\rm B}T/\epsilon=0.772$) and a highly supercooled liquid ($k_{\rm B}T/\epsilon=0.267$).
Here the $\alpha$ relaxation time $\tau_\alpha$ is defined using the self-intermediate scattering function $F_{\rm s}(k_m,\tau_\alpha) = e^{-1}$.
Note that $D_{j}$ is discrete in real space, and thus the plotted values are obtained through local spatial averaging.
A darker (or lighter) color corresponds to the regions with less (or more) mobile particles; hence, an increase in the color variation indicates that more significant DH appears in the system.
Accordingly, from these examples, DH becomes more intense at lower temperatures, {\it i.e.}, when the system is highly supercooled.

Based on the order parameter $Q_k$, we define the four-body correlation function as
\begin{equation}
    S_{4,k}(q,\tau) = \left<{ \hat{Q}_k(\boldsymbol{q},t,\tau) \hat{Q}_k(-\boldsymbol{q},t,\tau) }\right> _{t,q},
    \label{eq: 4point_corrfunc}
\end{equation}
where $\langle{\cdots}\rangle_{t,q}$ is the average over time $t$ and the angular components of $\bm{q}$.
The function $S_{4,k}$ represents the spatial correlation of the mobility of each particle in the time interval $\tau$.
The correlation length $\xi_{4,k}(\tau)$ and the intensity $\chi_{4,k}(\tau)$ are obtained by fitting $S_{4,k}(q,\tau)$ to the Orstein--Zernike (OZ) form
\begin{equation}
    S_{4,k}(q,\tau) = \frac {\chi_{4,k}(\tau)} {1 + q^2 \xi_{4,k}^2(\tau)}
    \label{eq: OZ}
\end{equation}
at small wavenumbers $q$ \cite{lavcevic2003, yamamoto1998}.
The values $\xi_{4,k}$ and $\chi_{4,k}$ thus represent the characteristic size and the intensity of DH, respectively.

\begin{figure}[tbp] 
    \centering
    \includegraphics[clip, width=1.0\linewidth]{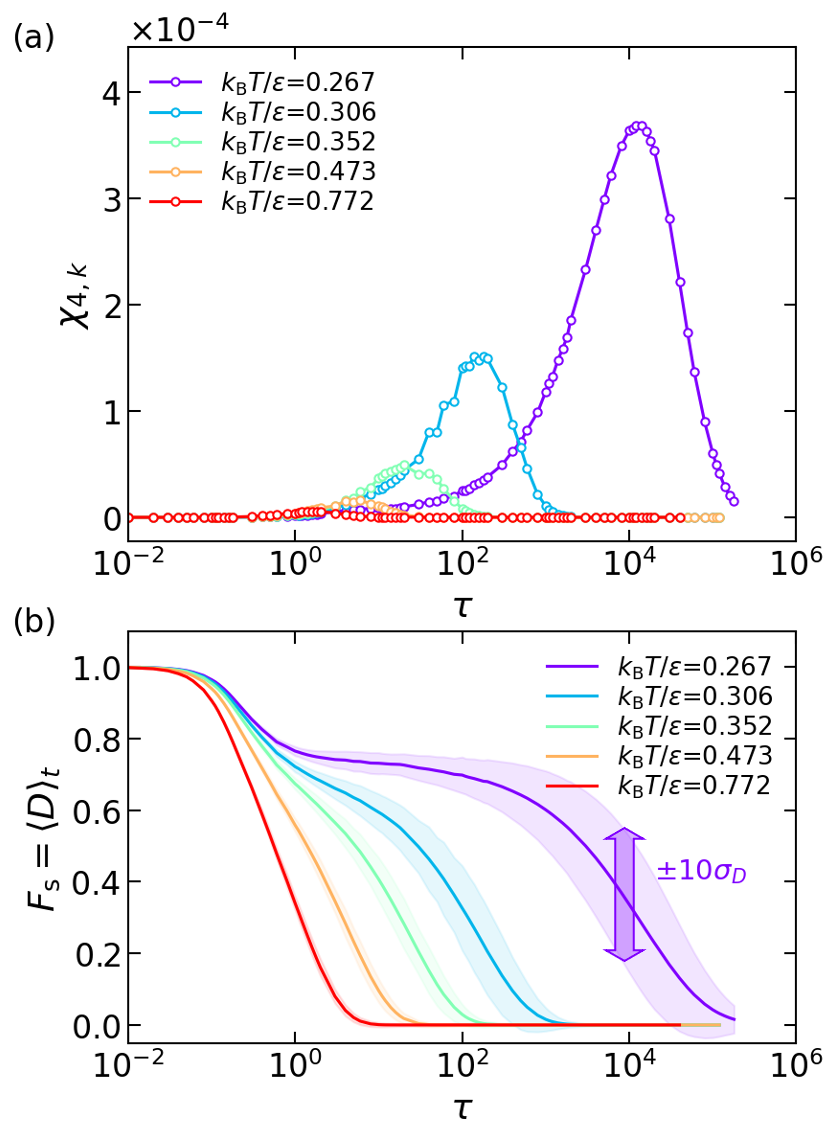}
\caption{
(a) The DH intensity $\chi_{4,k}$ as a function of the time interval $\tau$ at different temperatures ($k_{\rm B}T/\epsilon=0.267$, $0.306$, $0.352$, $0.473$, and $0.772$).
As the temperature decreases, the peak intensity, $\chi_{4,k}^*$, becomes more prominent, and the time interval when the peak occurs, $\tau_{4,k}^*$, is prolonged.
(b) The particle immobility averaged over all particles in the system and time $t$, $\langle D \rangle_t=F_{\rm s}(k_m,\tau)$.
Ten times the standard deviation at each time interval $\pm10\sigma_D$ is shown with the light colored areas around the average immobility $\langle D \rangle_t$.
}
    \label{fig: DH_4pointcorr}
\end{figure}

We have quantified $\chi_{4,k}$ by calculating the four-body correlation function $S_{4,k}$ using the method described in the previous paragraphs, and we plot the results against the time interval $\tau$ in Fig.~\ref{fig: DH_4pointcorr}~(a) at
different temperatures $k_{\rm B}T/\epsilon = 0.267$, $0.306$, $0.352$, $0.473$, and $0.772$.
This figure shows that the DH intensity $\chi_{4,k}$ reaches a peak $\chi_{4,k}^*$, which increases in height with decreasing temperature $T$.
Therefore, the dynamics of the system become more heterogeneous when it is more supercooled.
In addition, we show the dependence of the particle immobility averaged over all particles and time, $\langle{D}\rangle_t = \langle{ \frac{1}{N} \sum_{j=1}^N{D_j}}\rangle_t$, which is identical to the self-intermediate scattering function $F_{\rm s}(k_m,\tau)$, in Fig.\ref{fig: DH_4pointcorr} (b).
As the temperature decreases, $F_{\rm s}(k_m,\tau)$ exhibits drastic slowing with a more stretched form, consistent with previous studies \cite{yamamoto1998, miyazaki2004}.
In the same figure, the standard deviation $\sigma_D=\sqrt{\langle D^2\rangle_t-\langle D\rangle^2_t}$ is shown with the light colored areas.
The relaxation time $\tau_{\alpha}$ clearly corresponds to the time interval when the DH intensity peaks, $\tau_{4,k}^*$.

\begin{figure}[tbp] 
    \centering
    \includegraphics[clip, width=1.0\linewidth]{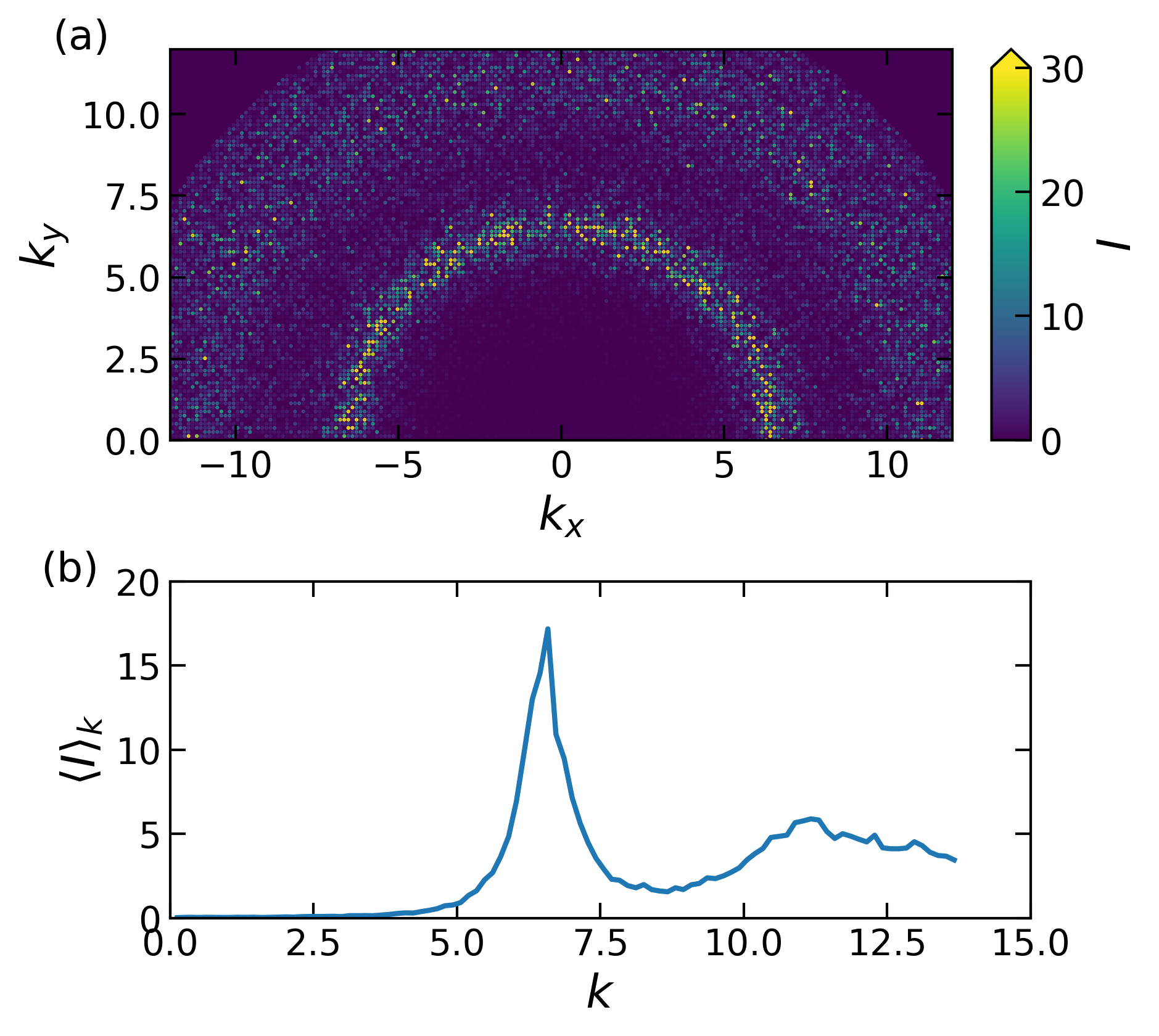}
\caption{
(a) The speckle pattern at a temperature of $k_{\rm B}T/\epsilon=0.267$.
The scattering intensity $I({\bm k}, t)$ in the speckle pattern reaches a peak on a circle at approximately $|{\bm k}|=k_m=2\pi$.
(b) The scattering intensity averaged over $\bm{k}$ vectors of the same magnitude, $\langle{I(\bm{k},t)}\rangle_{k}$.
The peak is confirmed in this figure.
}
    \label{fig: speckle_pattern}
\end{figure}

In previous experimental studies \cite{mayer2004, duri2005, berthier2011, hoshino2020}, the DH intensity was reportedly quantified based on speckle patterns measured in scattering experiments, such as XPCS.
The method discussed in the previous paragraph
cannot be applied to these experiments in practice; therefore, DH must be quantified from speckle patterns as an alternative.
While the scattering intensity $I$ is feasible to measure experimentally, we calculate it in our MD simulations using the following equation:
\begin{equation}
    I(\bm{k},t) = \rho_{\bm{k}}(t)\rho_{-\bm{k}}(t),
    \label{eq: intensity}
\end{equation}
where $\rho_{\bm{k}}(t)$ is the Fourier transform of the particle density $\rho(t) = \sum^N_{j=1}\sigma_j^3 \delta\left[{\bm{r}-\bm{r}_j(t)}\right]$ and $\bm{k}$ is a wave vector.
Fig.~\ref{fig: speckle_pattern}~(a) illustrates the scattering intensity in a speckle pattern calculated using Eq.~(\ref{eq: intensity}).
The pattern is plotted on a 2D plane with $k_z=0$ at $k_{\rm B}T/\epsilon=0.267$.
The scattering intensity $I$ varies with the vector $\bm{k}$, and it reaches a sharp peak with a magnitude of approximately $|{\bm k}|=k_m=2\pi$.
One can confirm this peak in Fig.~\ref{fig: speckle_pattern}~(b), where the intensity averaged over the angular components of $\bm{k}$, 
$\langle{I(\bm{k},t)}\rangle_{k}$, is plotted.

Using the scattering intensity $I(\bm{k},t)$, we calculate the correlation function between the speckle patterns at two different times using the following equation \cite{brown1997, malik1998}:
\begin{equation}
    C_I=C_I(k_m,t,\tau) = \frac{ \left<{I(\boldsymbol{k},t) I(\boldsymbol{k},t+\tau)}\right>_{k_m}}{ \left<{I(\boldsymbol{k},t)}\right>_{k_m} \left<{I(\boldsymbol{k},t+\tau)}\right>_{k_m} }.
    \label{eq: two-time_corrfunc}
\end{equation}
This function $C_I$ represents the relaxation process of the scattering intensity at time $t$.
In the presence of DH, $C_I$ fluctuates with time $t$, and thus the DH intensity is quantified by calculating the variance of this relaxation function \cite{duri2005, hoshino2020}.
Then, the normalized variance is defined as follows:
\begin{equation}
    \chi_k(\tau) = 
    \frac
    { \left<C^2_{I}\right>_t - \left<{C_{I}}\right>^2_{t} }
    { \left<{C_{I}(k_m,t,\tau=0)}\right>^2_{t} }
    \simeq
    \frac
    { \left<C^2_{I}\right>_t - \left<{C_{I}}\right>^2_{t} }
    {4 }.
    \label{eq: DH_speckle}
\end{equation}
The value $\chi_k(\tau)$ represents the intensity of DH at time interval $\tau$ and indicates the extent to which the particle dynamics vary locally.

\begin{figure}[tbp] 
    \centering
    \includegraphics[clip, width=1.0\linewidth]{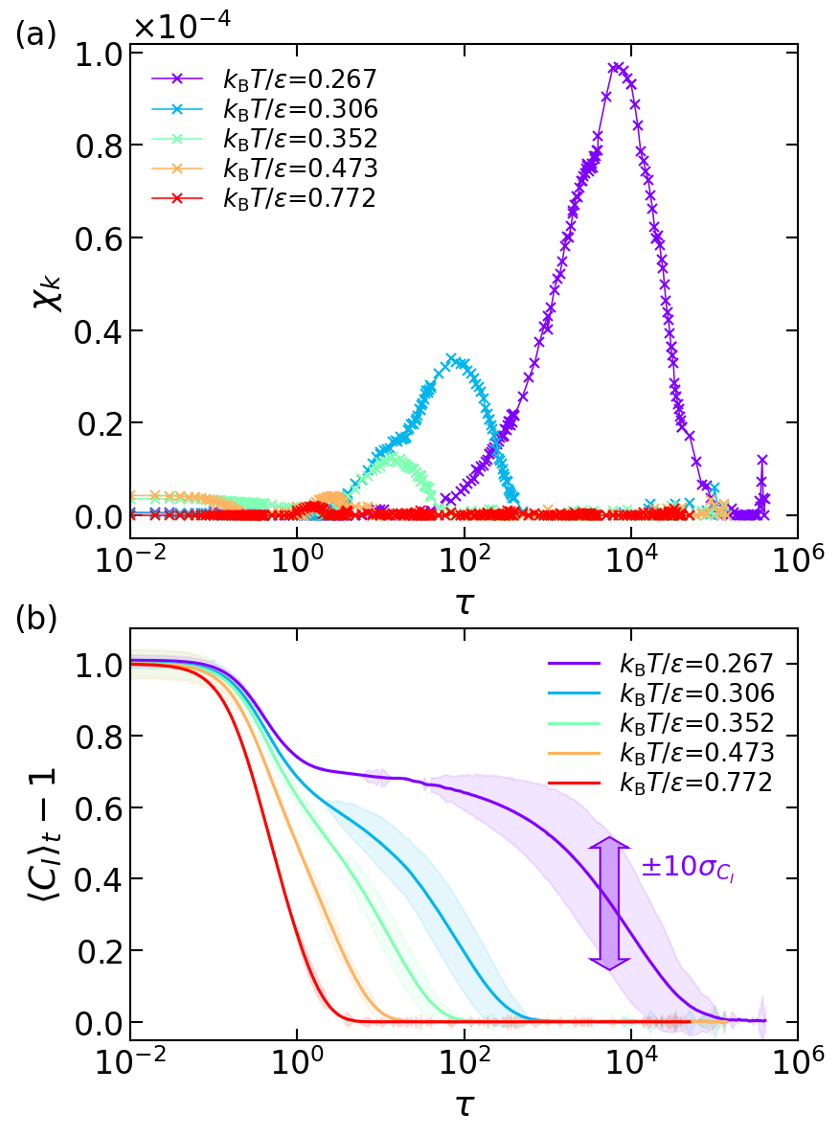}
\caption{
(a) The DH intensity $\chi_k$ as a function of the time interval $\tau$ at different temperatures ($k_{\rm B}T/\epsilon=0.267$, $0.306$, $0.352$, $0.473$, and $0.772$).
As the temperature decreases, the intensity peak, $\chi_k^*$, becomes more prominent, and the time interval when the peak occurs, $\tau^*$, becomes later.
(b) The two-time correlation function averaged over time $t$ with the same time interval $\tau$, $\langle{C_I}\rangle_t - 1$.
Ten times the standard deviation at each time interval $\pm10\sigma_{C_I}$ is shown with the light colored areas around the average immobility $\langle C_I \rangle_t-1$.
}
    \label{fig: DH_speckle}
\end{figure}

We have quantified the DH intensity $\chi_{k}(\tau)$ based on the scattering intensity $I({\bm k},t)$ as described above, and the results are shown in Fig.~\ref{fig: DH_speckle}~(a).
We have investigated the values at different temperatures $k_{\rm B}T/\epsilon=0.267$, $0.306$, $0.352$, $0.473$, and $0.772$.
Note that the value $\chi_{k}$ directly calculated from the equation contains some statistical noise due to the limited number of sampling points on the speckle pattern, $n_{\bm{k}}$.
We have applied a correlation procedure to the values by extrapolation to the case of $1/n_{\bm{k}}\rightarrow0$ ($n_{\bm{k}}\rightarrow\infty$) to overcome this problem of spatial resolution \cite{duri2005, trappe2007, hoshino2020}.
The values approximated using the method described above are plotted in the figure.
$\chi_k(\tau)$ reaches a peak at each temperature, and the height of this peak becomes more prominent in a more supercooled state.
These results are similar to those obtained using the approach based on the four-body correlation function and in previous experimental studies \cite{duri2006, hoshino2020, trappe2007}.
We also show the two-time correlation function calculated using Eq.~(\ref{eq: two-time_corrfunc}) and averaged over time $t$, $\langle{C_I}\rangle_t - 1$, in Fig.~\ref{fig: DH_speckle}~(b).
This averaged correlation represents the relaxation process of the scattering intensity $I$.
Similar features are observed when comparing these results and the intermediate scattering functions shown in Fig.~\ref{fig: DH_4pointcorr}~(b).
We also show the standard deviation $\sigma_{C_I}=\sqrt{\langle C_I^2\rangle_t-\langle C_I\rangle^2_t}$ with the light colored areas for each temperature.
Note that the values are multiplied to make the standard deviations easier to visualize.
As shown in this figure, DH becomes most significant around the relaxation time $\tau^*$ of $\langle{C_I}\rangle_t - 1$, which is almost equivalent to the $\alpha$ relaxation time $\tau_\alpha$ in Fig.=\ref{fig: DH_4pointcorr}.
Additionally, the peak intensity $\chi_k^*=\chi_k(\tau^*)$ increases when the system is in a more supercooled state, as also shown in Fig.~\ref{fig: DH_speckle}~(a).

Finally, we discuss the relation between the two approaches for evaluating the DH presented above.
According to previous experimental studies \cite{duri2005, hoshino2020, trappe2007}, the temporal two-time correlation function at $t$, ${C_I} - 1$, is approximated as a stretched exponential $\exp{[{ -\left[{ \gamma(t)\tau }\right]^{\mu(t)} }]}$, and the variance of $C_I(k_m,t,\tau)$ arises from the fluctuations of the two dynamic parameters, $\gamma(t)$ and $\mu(t)$.
Additionally, the temporal self-intermediate scattering function $D(k_m,t,\tau)$ has also been approximated in a similar form \cite{yamamoto1998}.
Therefore, we assume that both the mean value and the variance of $C_I-1$ are approximately equal to those of $D$, {\it i.e.}, 
\begin{equation}
        \left<{C_{I}}\right>_{t} - 1 \simeq \left<D\right>_{t}=F_s(k_m,\tau) \ \ {\rm and}\ \ 
        \sigma_{C_{I}}^2 \simeq  \sigma_D^2.
    \label{eq: assumption}
\end{equation}
Comparing Fig.~\ref{fig: DH_4pointcorr}~(b) and Fig.~\ref{fig: DH_speckle}~(b), the present simulation data support the validity of these assumptions.
From Eqs.~(\ref{eq: order_param_FT}), (\ref{eq: particle_mobility}), (\ref{eq: 4point_corrfunc}) and (\ref{eq: OZ}), the DH intensity is approximated as 
\begin{equation}
\chi_{4,k}(\tau) = \lim_{q\rightarrow0}S_{4,k} \simeq \left<D^2\right>_t -\left<D\right>^2_t .
    \label{eq: chi4_approx}
\end{equation}
From Eqs.~(\ref{eq: DH_speckle}),~(\ref{eq: assumption})~and~(\ref{eq: chi4_approx}), we can relate the two types of DH intensity obtained from the four-body correlation function and from the scattering intensity as follows:
\begin{equation}
    \chi_{4,k}(\tau)
    \approx 
    4\chi_{k}(\tau).
    \label{eq: DH_relation}
\end{equation}
As summarized in the Supplemental Material, the same result derived from Eq.~(\ref{eq: DH_relation}) is also obtained using an equation based on the Siegert relation \cite{siegert} with a slightly different choice for $D_j$ from Eq.~(\ref{eq: particle_mobility}).

\begin{figure}[tbp] 
    \centering
    \includegraphics[clip, width=1.0\linewidth]{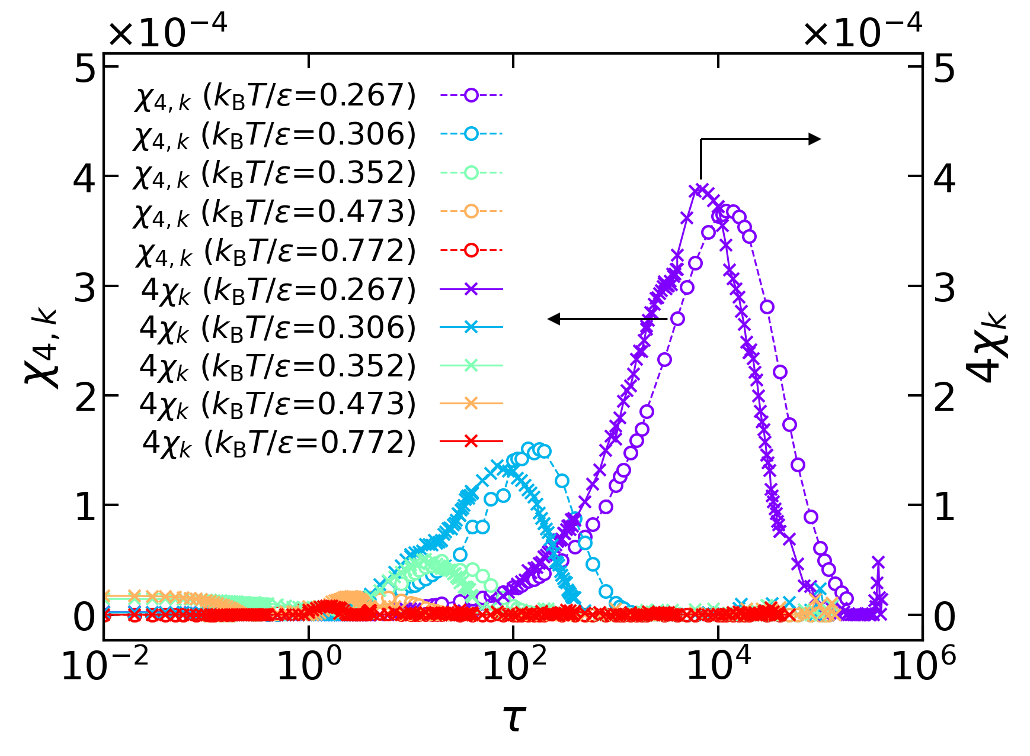}
\caption{
A comparison between the DH intensities obtained from the four-body correlation function, $\chi_{4,k}$, and those obtained from the speckle patterns, $4\chi_{k}$, as functions of the time interval $\tau$ at different temperatures ($k_{\rm B}T/\epsilon=0.267$, $0.306$, $0.352$, $0.473$, and $0.772$).
The two sets of results show good agreement with each other.
}
    \label{fig: DH_compare}
\end{figure}

We have checked the relation presented in Eq.~(\ref{eq: DH_relation}) by comparing the results obtained using both approaches, as shown in Fig.~{\ref{fig: DH_compare}}.
We show the dependence of $\chi_{4,k}$ and $4\chi_{k}$ on the time interval $\tau$, with different colors corresponding to different temperatures.
Note that $\chi_{4,k}$ is illustrated using circles, and $\chi_{k}$ is illustrated using crosses.
As shown in Fig.~{\ref{fig: DH_compare}}, the heights of the peaks, $\chi_{4,k}^*$ and $4\chi_{k}^*$, are approximately the same at each temperature.
Therefore, we conclude that Eq. (\ref{eq: DH_relation}) is valid to some extent and that the two approaches discussed above focus on the same physical property of the system.

In conclusion, in previous studies, only one of the following two routes has been used to quantitatively characterize DH in any particular case: 1) using four-body correlation functions or 2) using the speckle patterns observed in scattering experiments.
In the present study, we successfully computed the intensity of DH using both methods by analyzing the same simulation data and compared the results in detail as functions of the temperature and the separation time.
We confirmed a high level of agreement between the two sets of results throughout the whole parameter range of the present MD simulations.
The present findings provide strong evidence for physical consistency between the DHs characterized using routes 1) and 2).

\begin{acknowledgments}
This work was supported by Grants-in-Aid for Scientific Research (JSPS KAKENHI) under grant nos. JP 20H00129 and 20H05619.
\end{acknowledgments}

\bibliography{ref.bib}

\clearpage
\begin{widetext}
\setcounter{page}{1}

\begin{center}
Supplemental Material for the manuscript\\
``Relation between dynamic heterogeneities observed in scattering experiments and four-body correlations''\\
\ \\
Kohsei Kanayama$^1$, Taiki Hoshino$^2$, Ryoichi Yamamoto$^1$
\\
$^1$Department of Chemical Engineering, Kyoto University, Kyoto 615-8510, Japan\\
$^2$RIKEN SPring-8 Center, 1-1-1 Kouto, Sayo-cho, Sayo-gun, Hyogo 679-5148, Japan
\end{center}

\section{SM1. Assessment of Eq.~(\ref{eq: DH_relation}) using the Siegert relation}
We also discuss the relation between the two approaches for evaluating DH using the Siegert relation \cite{siegert}:
\begin{equation}
    g_2(k,\tau)
    =
    1 + |g_{1}(k,\tau)|^2,
    \label{eq: Siegert}
\end{equation}
where $g_2$ is the time correlation function of scattering intensity $I$ and $g_1$ is the normalized full-intermediate scattering function.
Considering 
\begin{equation}
g_2(k_m,\tau) = \langle C_{I} \rangle_t
\end{equation}
and 
\begin{equation}
g_{1}(k_m,\tau) = F(k_m,\tau)/F(k_m,\tau=0) \simeq F_{\rm s}(k_m,\tau) = \frac{1}{N}
\left<\left<\sum_{j=1}^{N}{ e^{-i\boldsymbol{k}\cdot \Delta\boldsymbol{r}_j(t,\tau)} }\right>_{k_m}\right>_t,
\end{equation}
Eq.(\ref{eq: Siegert}) leads to 
\begin{equation}
    \langle C_{I} \rangle_t -1
    =
\left[F_{\rm s}(k_m,\tau)\right]^2.
    \label{eq: Siegert2}
\end{equation}
Eq.(\ref{eq: Siegert2}) is apparently different from the empirically supported assumption, which we introduced as Eq.(\ref{eq: assumption}), but still holds quite well as we will see in Fig.\ref{fig: Sigert relation}.
Here, we have further assumed the instantaneous relation between the two functions as follows:
\begin{equation}
    C_{I}-1
    \simeq
    \frac{1}{N^2}\left<\sum_{j=1}^{N}{ e^{-i\boldsymbol{k}\cdot \Delta\boldsymbol{r}_j(t,\tau)} }\right>^2_{k_m}.
    \label{eq: Siegert assumption}
\end{equation}
Therefore, 
\begin{equation}
    \chi_k(\tau) 
    \simeq
    \frac
    { \left<C^2_{I}\right>_t - \left<{C_{I}}\right>^2_{t} }
    {4}
    =\frac{1}{4}\left[
    \left<
    \left(\frac{1}{N^2}\left<\sum_{j=1}^{N}{ e^{-i\boldsymbol{k}\cdot \Delta\boldsymbol{r}_j(t,\tau)} }\right>^2_{k_m}\right)^2
    \right>_t - \left<
    \frac{1}{N^2}\left<\sum_{j=1}^{N}{ e^{-i\boldsymbol{k}\cdot \Delta\boldsymbol{r}_j(t,\tau)} }\right>^2_{k_m}
    \right>^2_{t}
    \right].
    \label{eq: chi_k_Siegert}
\end{equation}

On the other hand, if the particle immobility in Eq. (\ref{eq: order_param}) is defined slightly different from Eq.~(\ref{eq: particle_mobility}) as
\begin{equation}
    D_j(k_m,t,\tau) = \left<{ e^{-i\boldsymbol{k}\cdot \Delta\boldsymbol{r}_j(t,\tau)} }\right>^2_{k_m},
    \label{eq: particle_mobility_squared}
\end{equation}
the DH intensity is obtained from the equation
\begin{equation}
    \chi_{4,k}(\tau)
    =\left<D^2\right>_t-\left<D\right>^2_t
    =\left<\left(\frac{1}{N}\sum_{j=1}^{N}\left<{ e^{-i\boldsymbol{k}\cdot \Delta\boldsymbol{r}_j(t,\tau)} }\right>^2_{k_m}\right)^2\right>_t-\left<\frac{1}{N}\sum_{j=1}^{N}\left<{ e^{-i\boldsymbol{k}\cdot \Delta\boldsymbol{r}_j(t,\tau)} }\right>^2_{k_m}\right>^2_t.
    \label{eq: chi4_approx_squared}
\end{equation}
Note that if we accept a crude order estimation
\begin{equation}
\frac{1}{N^2}\left<\sum_{j=1}^{N}{ e^{-i\boldsymbol{k}\cdot \Delta\boldsymbol{r}_j(t,\tau)} }\right>^2_{k_m}
\sim
\frac{1}{N}\sum_{j=1}^{N}\left<{ e^{-i\boldsymbol{k}\cdot \Delta\boldsymbol{r}_j(t,\tau)} }\right>^2_{k_m}
\label{eq:crude}
\end{equation}
assuming no correlations exist between $\Delta\boldsymbol{r}_i$ and $\Delta\boldsymbol{r}_j$ for $i\neq j$
and use Eqs.~(\ref{eq: chi_k_Siegert})~and~(\ref{eq: chi4_approx_squared}), the two types of DH intensity obtained from the four-body correlation function and from the scattering intensity are again related as follows:
\begin{equation}
    \chi_{4,k}(\tau)
    \approx
    4\chi_{k}(\tau),
    \label{eq: DH_relation_Siegert}
\end{equation}
which is equivalent to Eq.~(\ref{eq: DH_relation}).

To test the validity of the above discussion numerically, we have assessed whether the Siegert relation is valid for the current system.
We show the comparison between the time correlation function of scattering intensity $ \langle C_{I} \rangle_t - 1$ and the squared intermediate scattering function $|g_1|^2$ in Fig.~\ref{fig: Sigert relation}.
The two functions show good agreement with each other, and Eq. (\ref{eq: Siegert2}) works in the MD simulations.
\begin{figure}[tbp] 
    \centering
    \includegraphics[clip, width=0.7\linewidth]{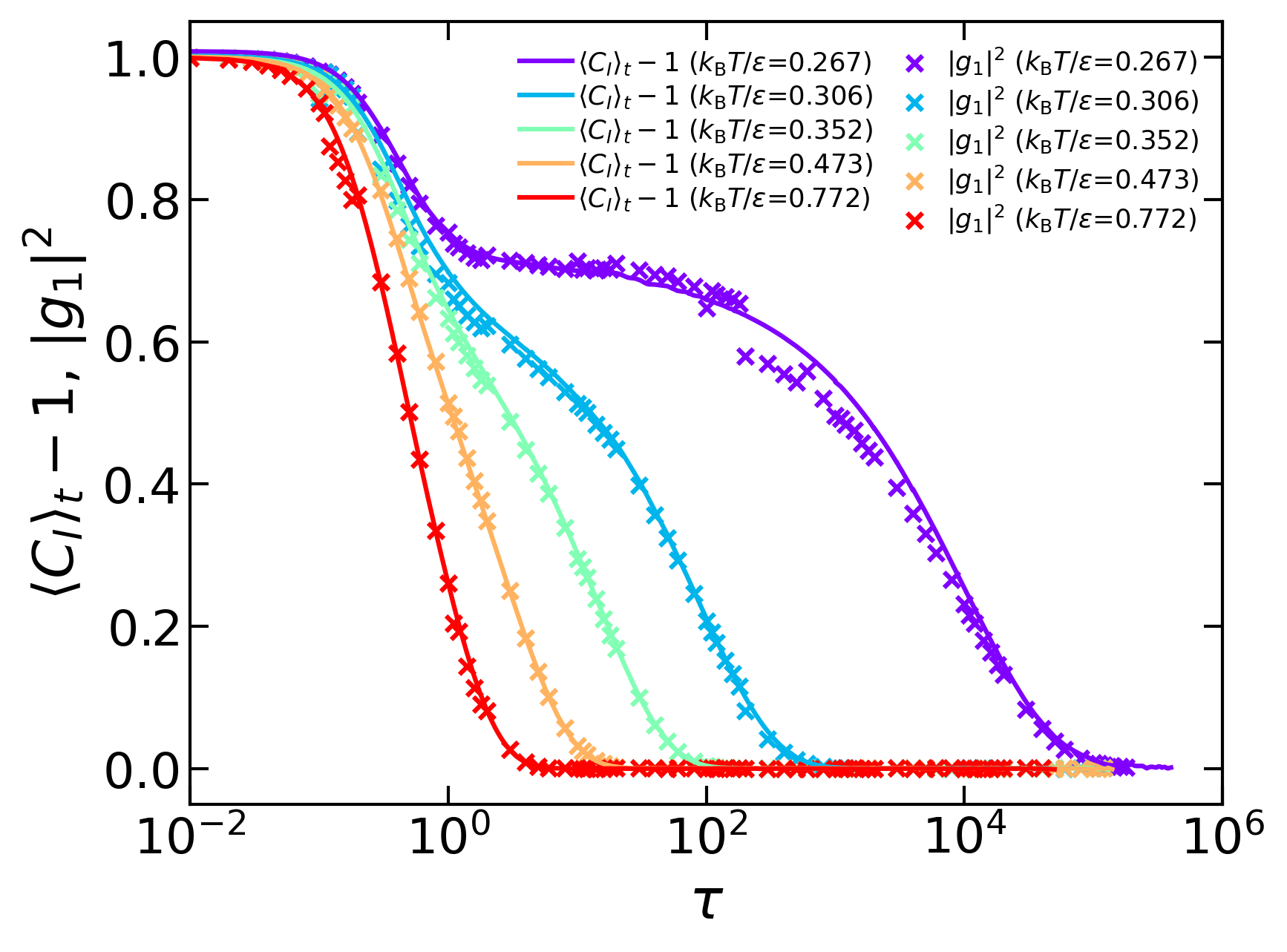}
    \caption{
    A comparison between the time correlation function of scattering intensity $ \langle C_{I}(k_m,t,\tau) \rangle_t - 1$ and the squared intermediate scattering function $|g_1(k_m,\tau)|^2$, as functions of the time interval $\tau$ at different temperatures ($k_{\rm B}T/\epsilon=0.267$, $0.306$, $0.352$, $0.473$, and $0.772$).
    The two functions are consistent with each other, and the Siegert relation (\ref{eq: Siegert}) is confirmed.
    }
    \label{fig: Sigert relation}
\end{figure}
We also have assessed the relation presented in Eq. (\ref{eq: DH_relation_Siegert}) by comparing the results obtained using both approaches, as shown in Fig.~{\ref{fig: DH_compare_Siegert}}.
We show the dependence of $\chi_{4,k}$ and $4\chi_{k}$ on the time interval $\tau$, with different colors corresponding to different temperatures.
Note that $\chi_{4,k}$ is illustrated using circles, and $\chi_{k}$ is illustrated using crosses.
As shown in Fig.~{\ref{fig: DH_compare_Siegert}}, the heights of the peaks, $\chi_{4,k}^*$ and $4\chi_{k}^*$, are approximately the same at each temperature.
However, the gaps between the two DH intensities in Fig.~\ref{fig: DH_compare} are smaller than those in Fig.~\ref{fig: DH_compare_Siegert}, while the gaps in Fig.~\ref{fig: DH_compare_Siegert} can be reduced if an adjustable parameter is introduced in the crude estimation Eq.(\ref{eq:crude}).
\begin{figure}[tbp] 
    \centering
    \includegraphics[clip, width=0.7\linewidth]{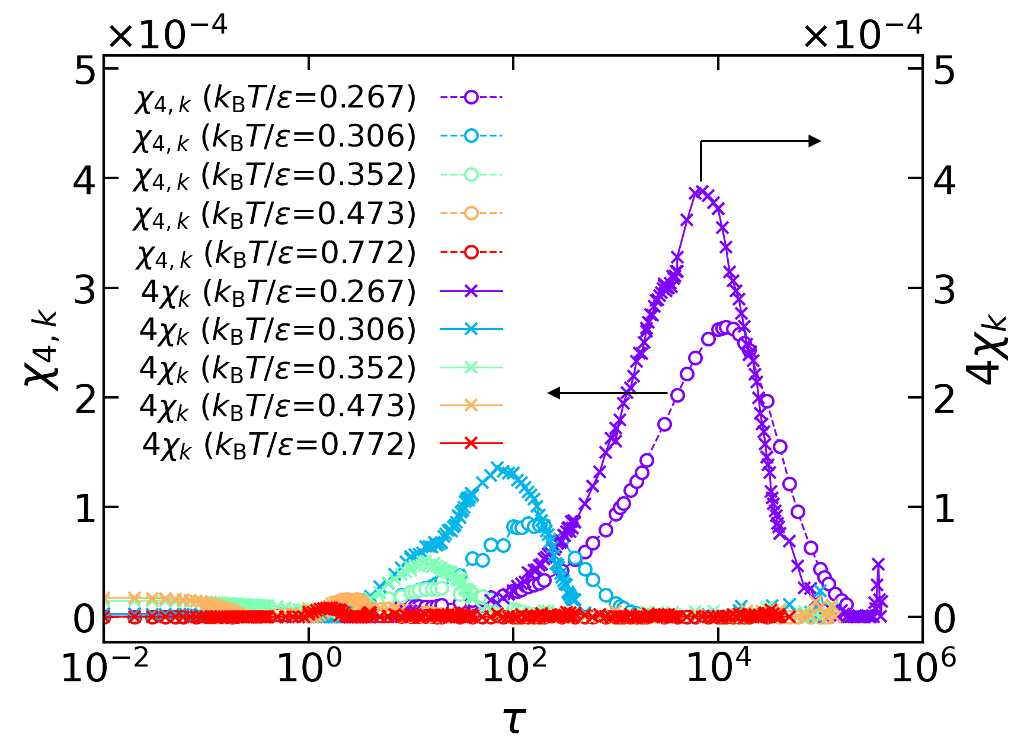}
    \caption{
    A comparison between the DH intensities obtained from the four-body correlation function, $\chi_{4,k}$, and those obtained from the speckle patterns, $\chi_{k}$, as functions of the time interval $\tau$ at different temperatures ($k_{\rm B}T/\epsilon=0.267$, $0.306$, $0.352$, $0.473$, and $0.772$).
    The two sets of results show good agreement with each other.
    }
    \label{fig: DH_compare_Siegert}
\end{figure}



\end{widetext}

\end{document}